\documentclass{iopconfser}

\usepackage{bm}
\usepackage[colorlinks=true]{hyperref}
\usepackage{physics}
\usepackage{mathtools}
\usepackage{tikz-feynhand}
\usepackage{ascmac}
\usepackage{cite}
\usepackage{graphics}
\usepackage{ragged2e}

\hypersetup{
    colorlinks=true,
    citecolor=blue,
    linkcolor=black,
    urlcolor=black,
}

\begin{document}

\title{Spin-lattice relaxation for point-node--like $\bm{s}$-wave superconductivity in $f$-electron systems}

\author{Shingo Haruna$^{1*}$, Koki Doi$^{1}$, Takuji Nomura$^{1, 2}$ and Hirono Kaneyasu$^{3}$}

\affil{$^1$Department of Material Science, University of Hyogo, 3-2-1 Kouto, Ako, Hyogo 678-1297, Japan}
\affil{$^2$Synchrotron Radiation Research Center, National Institutes for Quantum Science and Technology, 1-1-1 Kouto, Sayo, Hyogo 679-5148, Japan}
\affil{$^3$Center for Liberal Arts and Sciences, Sanyo-Onoda City University, 1-1-1 Daigakudori, Sanyo-Onoda, Yamaguchi 756-0884, Japan}

\email{$^*$rk24y005@guh.u-hyogo.ac.jp}

\begin{abstract}
In this study, we examined the temperature dependence of the spin-lattice relaxation using an $f$-$d$-$p$ model, which is an effective model of UTe$_2$.
Solving the linearized Eliashberg equation in the $f$-$d$-$p$ model based on third-order perturbation theory, we obtain a point-node--like $s$-wave pairing state.
Our result shows that the Hebel-Slichter-peak in the point-node--like $s$-wave pairing state is smaller than that in the isotropic $s$-wave pairing state.
However, the Hebel-Slichter peak remains robust even in the point-node--like $s$-wave pairing state, and the point-node--like $s$-wave state is inconsistent with the results of nuclear magnetic resonance measurements.
\end{abstract}

\begin{justify}
\section{Introduction}
UTe$_2$ has attracted attention due to its interesting superconducting properties \cite{Ran2019,Aoki2022}, for instance, a large upper critical field exceeding the Pauli-limiting field \cite{Ran2019,Aoki2022,Knebel2019,Sakai2023,Rousel2023}, the multiple superconducting phases under pressure and field \cite{Aoki2022,Knebel2019,Sakai2023,Rousel2023,Braithwaite2019,Aoki2020,Lin2020,Ran2020,Aoki2021}, a point-node--like behavior of the specific heat \cite{Aoki2022,Metz2019,Kittaka2020,Rosa2022,Sakai2022,Weiland2022,Ishihara2023}, and so on.
UTe$_2$ exhibits a critical temperature $T_c$ of up to 2.1 K and this compound is expected as a candidate of the spin-triplet superconductor because its upper critical field exceeds the Pauli-limiting field in all crystal directions.
Recent nuclear magnetic resonance (NMR) measurements showed that the NMR Knight shift decreases in all crystalline directions \cite{Matsumura2023}.
This result indicates that the pairng state in UTe$_2$ is the spin-singlet states or the spin-triplet $\mathrm{A_u}$ state.
The spin-triplet $\mathrm{A_u}$ state usually shows a fully gapped superconducting state, however, specific heat \cite{Aoki2022,Metz2019,Kittaka2020,Rosa2022,Sakai2022,Weiland2022,Ishihara2023} and penetration depth \cite{Ishihara2023} measurements indicate the possibility of the existence of the point-node.
In addition, the penetration depth measurements suggest that it is difficult to represent the positions of the nodes using a single irreducible representation \cite{Ishihara2023}.
The irreducible representation of the superconducting state in UTe$_2$ remains an open question, and determining the node structure of the superconducting gap in UTe$_2$ is an important issue.
Moreover, recent neutron scattering measurements have observed not ferromagnetic fluctuations but antiferromagnetic fluctuations with $\bm{Q} \approx (0, \pm \pi, 0)$ \cite{Duan2021,Butch2022}.
Although early studies considered that the ferromagnetic fluctuations lead to the spin-triplet pairing state in UTe$_2$, these observations imply the possibility that superconducting state may arise from correlations other than the ferromagnetic fluctuations.

In our previous study, to investigate an effect of the interaction that is not mediated by magnetic fluctuations on the superconductivity in UTe$_2$, we solved the linearized Eliashberg equation within the third-order perturbation theory (TOPT) \cite{Hotta1994,Nomura2000,Nishikawa2002,Kaneyasu2002,Kaneyasu2003-1,Kaneyasu2003-2,Yanase2005,Kaneyasu2005,Nomura2009} and obtained the $s$-wave pairing state, which has a point-node--like gap structure, as the most probable pairing state \cite{Haruna2024}.
Since this $s$-wave pairing state with accidental nodes is mediated by the on-site Coulomb repulsion, it is unconventional superconductivity.
This point-node--like gap structure leads to a $T^3$ behavior of the specific heat at low temperatures in alignment with the specific heat measurements \cite{Aoki2022,Metz2019,Kittaka2020,Rosa2022,Sakai2022,Weiland2022,Ishihara2023}.
Such an $s$-wave pairing state with accidental node structure, which is not symmetry protected, can account for both onserved the reduction in the NMR Knight shift in all crystalline directions and the point-node--like behavior of the specific heat.
In addition, we introduced the “$f$-$d$-$p$ model” as an effective model of UTe$_2$ \cite{Haruna2024} that reproduced the following observations: quasi-two-dimensional Fermi surfaces \cite{Fujimori2019,Miao2020,Aoki2023} and a magnetic fluctuation peak at $\bm{Q} \approx (0, \pm \pi, 0)$ \cite{Duan2021,Butch2022}.

Nuclear magnetic resonance measurements in UTe$_2$ reported the absence of a peak structure in the spin-lattice relaxation rate just below $T_c$ \cite{Nakamine2019,Matsumura2023S}.
In general, a large coherence peak is realized in the density of state (DOS) in the isotropic $s$-wave superconductivity, and this coherence peak encourages low-energy excitation just below $T_c$ and leads to the peak structure in the NMR spin-lattice relaxation $1/T_1$, the so-called “Hebel-Slichter peak” \cite{Hebel1959a,Hebel1959b}.
In this work, we investigate the temperature dependence of the $1/T_1$ in the point-node--like $s$-wave superconductivity obtained by solving the linearized Eliashberg equation within TOPT to determine whether the point-node--like $s$-wave pairing is consistent with NMR measurements.
Since the gap anisotropy leads to a broadening of the peak structure of the DOS, we considered the possibility that this broadening is the reason that the Hebel-Slichter peak is not observed in UTe$_2$.
To this end, we calculate the site-resolved temperature dependency of $1/T_1$ and partial density of state (PDOS) in the superconducting phase using the point-node--like gap structure given by the linearized Eliashberg equation.
Our calculation result shows that the Hebel-Slichter peak remains robust even in the highly anisotropic superconductivity and the point-node--like $s$-wave pairing state is inconsistent with the results of NMR \cite{Nakamine2019,Matsumura2023S}.

\section{$f$-$d$-$p$ model and gap function of anisotropic $s$-wave SC}
The Hamiltonian of the $f$-$d$-$p$ model introduced in Ref. \cite{Haruna2024} is given by $\hat{H} = \hat{H}_0 + \hat{H}_I$, where $\hat{H}_0$ and $\hat{H}_I$ are a six-orbital tight-binding Hamiltonian and the on-site Coulomb interaction between $f$ electrons, respectively.
In this tight-binding model, six orbitals, namely $f, f', d, d', p$ and $p'$, are taken into consideration, and the details of this model are available in Ref. \cite{Haruna2024S}.
By diagonalizing $\hat{H}_0$, we can obtain the band dispersion $\xi_a(\bm{k})$ and the unitary matrix $U_{la}(\bm{k})$, where indices $a$ and $l$ represent the band and orbital states, respectively.
As discussed in our previous study \cite{Haruna2024}, $f$-$d$-$p$ model reproduces the two-dimensional Fermi surfaces and anti-ferromagnetic fluctuations with $\bm{Q} \approx (0, \pi, 0)$.
The reproduced Fermi surfaces consist of $\alpha$- and $\beta$-surfaces, which are the hole and electron surfaces, respectively.

In our previous study, we solved the linearized Eliashberg equation
\begin{align}
\lambda \Delta_a(k) = - \frac{T}{N} \sum \limits_{k'} \sum \limits_{a'} V_{aa'}^{S/T}(k, k') \left| \mathcal{G}^{(0)}_{a'}(k') \right|^2 \Delta_{a'}(k')
\label{eb}
\end{align}
using the $f$-$d$-$p$ model, where $\lambda$ and $\Delta_a(k)$ are the eigenvalue and anomalous self-energy for the band $a$ state, respectively.
$\mathcal{G}^{(0)}_a(k)$ is the bare thermal Green's function given by $\left( i \omega_n - \xi_a(\bm{k}) \right)^{-1}$ using the band dispersion $\xi_a(\bm{k})$ and fermionic Matsubara frequencies $\omega_n = (2n + 1) \pi T$.
$V_{aa'}^{S/T}(k, k')$ is the effective pairing interaction, which represents the pair scattering amplitude from $(k'~a', -k'~a')$ to $(k~a, -k~a)$ for spin-singlet/triplet channels, and it is approximated within TOPT in our previous study \cite{Haruna2024}.
Our investigation showed that the most probable pairing state in the $f$-$d$-$p$ model within TOPT is the highly anisotropic $s$-wave pairing state.
This $s$-wave state has a point-node--like structure at corners of the $\alpha$-surface in $k_z = 0$ plane, and it is an accidental node structure.
Such node structure allows zero-energy excitations and leads to the $T^3$ behavior of the specific heat although in $s$-wave superconductivity \cite{Haruna2024}.

\section{Spin-lattice relaxation in superconductivity}
\begin{figure}[t]
\centering
\includegraphics[width = 0.75\linewidth]{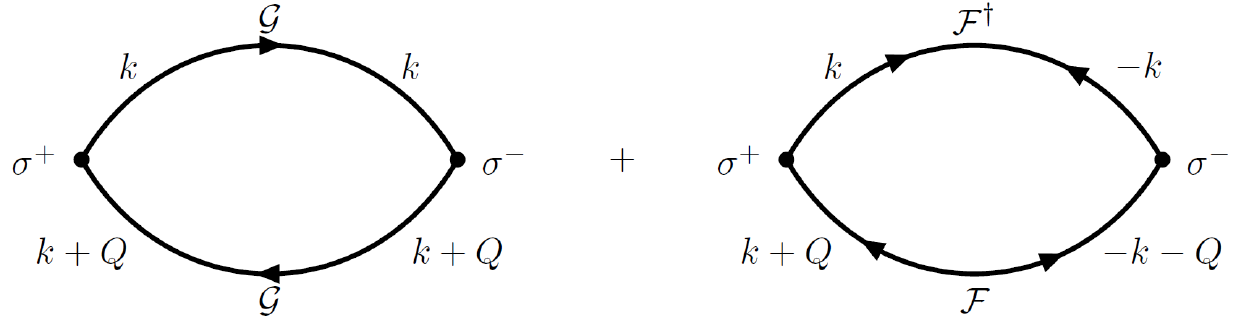}
\caption{Diagrammatic representation of the transverse spin susceptibility in a superconducting state.
$\sigma^+ \equiv \sigma^x + i \sigma^y$ and $\sigma^- \equiv \sigma^x - i \sigma^y$ are ladder operators of the spin state.}
\label{fig1}
\end{figure}
The spin-lattice relaxation rate in NMR measurements $1/T_1$ is given by 
\begin{align}
\frac{1}{T_1T} \propto \lim_{\Omega \to 0} \sum \limits_{\bm{Q}} \frac{\mathrm{Im} \chi^R_{+-}(\bm{Q}, \Omega)}{\Omega},
\label{imchi}
\end{align}
where $\chi^R_{+-}(\bm{Q}, \Omega)$ is the retarded spin susceptibility component perpendicular to the $z$ axis \cite{Moriya1963}.
This retarded spin susceptibility is obtained by the analytic continuation of the thermal spin susceptibility $\chi_{+-}(\bm{Q}, i \Omega_m)$ which is given by
\begin{align}
\chi_{+-}(Q)  = - \frac{T}{N} \sum \limits_k \sum \limits_{l_1 l_2}
\Big[
\mathcal{G}_{l_2 l_1}(k) \mathcal{G}_{l_1 l_2}(k + Q) +
\mathcal{F}^\dagger_{l_1 l_2}(k) \mathcal{F}_{l_1 l_2}(k + Q)
\Big]
\label{chi}
\end{align}
in the spin-singlet superconducting condition, where the thermal Green's functions are given by
\begin{align}
\mathcal{G}_{l l'}(k) &= - \int_0^\beta \ev{T_\tau \Big[c_{\bm{k} l \uparrow}(\tau)c^\dagger_{\bm{k} l' \uparrow} \Big]} e^{i \omega_n \tau} d\tau, \\
\mathcal{F}_{l l'}(k) &= \int_0^\beta \ev{T_\tau \Big[c_{\bm{k} l \uparrow}(\tau) c_{-\bm{k} l' \downarrow} \Big]} e^{i \omega_n \tau} d\tau, \\
\mathcal{F}^\dagger_{l l'}(k) &= \int_0^\beta \ev{T_\tau \Big[c^\dagger_{-\bm{k} l' \downarrow}(\tau) c^\dagger_{\bm{k} l \uparrow} \Big]} e^{i \omega_n \tau} d\tau.
\end{align}
This thermal spin susceptibility $\chi_{+-}(Q)$ is shown diagrammatically in Fig. \ref{fig1}.
The four-dimensional momenta $k$ and $Q$ are defined by $k \equiv (\bm{k}, i \omega_n)$ and $Q \equiv (\bm{Q}, i \Omega_m)$, where $\omega_n = (2n+1) \pi T$ and $\Omega_m = 2m \pi T$ are the fermionic and bosonic Matsubara frequencies, respectively.
The indices $l_1$ and $l_2$, which indicate the orbital states in the tight-binding model, are related to the band state $a$ using the unitary matrix $U_{la}(\bm{k})$ obtained by diagonalizing the tight-binding Hamiltonian.
Hereinafter, we assume that $U_{la}(\bm{k}) = U^*_{la}(-\bm{k})$.
The analytic continuation of the thermal Green's functions leads to the band diagonalized retarded Green's functions
\begin{align}
G^R_{a}(\bm{k}, \varepsilon) &= - \frac{\varepsilon + i \gamma + \xi_a(\bm{k})}{E^2_a(\bm{k}) - (\varepsilon + i\gamma)^2}, 
\label{gf} \\
F^R_{a}(\bm{k}, \varepsilon) &= \frac{\Delta_a(\bm{k})}{E^2_a(\bm{k}) - (\varepsilon + i\gamma)^2},
\label{agf}
\end{align}
where $E_a(\bm{k}) = \sqrt{\xi_a^2(\bm{k}) + |\Delta_a(\bm{k})|^2}$ \cite{Mineev1999}.
$\gamma$ is a small real value related to the lifetime of the quasi-particles.
$\xi_a(\bm{k})$ and $\Delta_a(\bm{k})$ are the energy dispersion and the superconducting gap function of the band state $a$, respectively.
The analytic continuation of Eq. (\ref{chi}) leads to the retarded susceptibility as follows:
\begin{align}
\chi^R_{+-}(\bm{Q}, \Omega) \propto \sum \limits_{\bm{k}} \sum \limits_{l_1 l_2} \int_{-\infty}^\infty d\varepsilon
&\bigg[
\tanh \frac{\varepsilon}{2T} \Big{\{}
G^R_{l_1 l_2}(\bm{k}+\bm{Q}, \varepsilon + \Omega) G^R_{l_2 l_1}(\bm{k}, \varepsilon)
\notag \\
&~~~~~~~~~~~~~~~ -
G^R_{l_1 l_2}(\bm{k}+\bm{Q}, \varepsilon + \Omega) G^A_{l_2 l_1}(\bm{k}, \varepsilon)
\notag \\
&~~~~~~~~~~~~~~~+
F^R_{l_1 l_2}(\bm{k}+\bm{Q}, \varepsilon + \Omega) F^R_{l_1 l_2}(\bm{k}, \varepsilon)
\notag \\
&~~~~~~~~~~~~~~~ -
F^R_{l_1 l_2}(\bm{k}+\bm{Q}, \varepsilon + \Omega) F^A_{l_1 l_2}(\bm{k}, \varepsilon)
\Big{\}}
\notag \\
&+
\tanh \frac{\varepsilon + \Omega}{2T} \Big{\{}
G^R_{l_1 l_2}(\bm{k}+\bm{Q}, \varepsilon + \Omega) G^A_{l_2 l_1}(\bm{k}, \varepsilon)
\notag \\
&~~~~~~~~~~~~~~~~~~~~ -
G^A_{l_1 l_2}(\bm{k}+\bm{Q}, \varepsilon + \Omega) G^A_{l_2 l_1}(\bm{k}, \varepsilon)
\notag \\
&~~~~~~~~~~~~~~~~~~~~ + 
F^R_{l_1 l_2}(\bm{k}+\bm{Q}, \varepsilon + \Omega) F^A_{l_1 l_2}(\bm{k}, \varepsilon)
\notag \\
&~~~~~~~~~~~~~~~~~~~~ -
F^A_{l_1 l_2}(\bm{k}+\bm{Q}, \varepsilon + \Omega) F^A_{l_1 l_2}(\bm{k}, \varepsilon)
\Big{\}}
\bigg],
\label{rchi}
\end{align}
where we assume that the gap function $\Delta_a(\bm{k})$ is a real function.
Defining two functions as
\begin{align}
N_{l_1 l_2}(\varepsilon) &\equiv \sum \limits_{\bm{k}} \sum \limits_{a} U_{l_1 a}(\bm{k}) U^*_{l_2 a}(\bm{k}) \mathrm{Im} G^R_a(\bm{k}, \varepsilon), \\
M_{l_1 l_2}(\varepsilon) &\equiv \sum \limits_{\bm{k}} \sum \limits_{a} U_{l_1 a}(\bm{k}) U^*_{l_2 a}(\bm{k}) \mathrm{Im} F^R_a(\bm{k}, \varepsilon),
\end{align}
and using Eq. (\ref{imchi}) and Eq. (\ref{rchi}), the spin-lattice relaxation rate is given by
\begin{align}
\frac{1}{T_1} &\propto
\int_{-\infty}^\infty d\varepsilon \; \frac{1}{\cosh^2[\varepsilon/2T]} 
\sum \limits_{l_1 l_2} \Big[
N_{l_1 l_2}(\varepsilon) N_{l_2 l_1}(\varepsilon)
+ M_{l_1 l_2}(\varepsilon) M_{l_2 l_1}(\varepsilon)
\Big].
\label{t1}
\end{align}
It is noted that this equation contains a summation with respect to $l_1$ and $l_2$.
By controlling the range of this summation, we can restrict orbitals that contribute to $1/T_1$.
In the following, the range of this summation is set to $l_1, l_2 = p, p'$ since actual NMR experiments in $f$-electron systems often use signals from light elements.

Since the temperature dependence of the superconducting gap $\Delta(T)$ is needed to calculate Eq. (\ref{t1}), thus, in this study, we assume that the superconducting gap is separatable into a momentum dependent part $g_a(\bm{k})$ and a temperature-dependent part $\Delta(T)$, that is, $\Delta_a(\bm{k}) = A_a g_a(\bm{k}) \Delta(T)$ \cite{Haruna2024,Nomura2002} where $A_a$ is a scaling factor.
The momentum-dependent part $g_a(\bm{k})$ has already been obtained by solving the linearized Eliashberg equation in Ref. \cite{Haruna2024}.
Based on this assumption, we can rewrite the effective pairing interaction $V_{aa'}(\bm{k}, \bm{k}')$ to $- A_a g_a(\bm{k}) A^*_{a'} g^*_{a'}(\bm{k}')$, therefore, the Bardeen--Cooper--Schrieffer (BCS) gap equation becomes
\begin{align}
1 = \frac{1}{N} \sum \limits_{\bm{k} a} \frac{\tanh \left[ E_a(\bm{k})/2T \right]}{E_{a}(\bm{k})} \left| A_a g_a(\bm{k}) \right|^2.
\label{bcs}
\end{align}
Solving Eq. (\ref{bcs}) numerically, we can estimate the temperature dependence of the superconducting gap magnitude $\Delta(T)$.
We note that the momentum dependency of the gap function $g_a(\bm{k})$ at $T = T_c$ is assumed to be unchanged to $T = 0$ in this formulation.

\section{Numerical results}
\begin{figure}[!b]
\centering
\includegraphics[width = 0.5\linewidth]{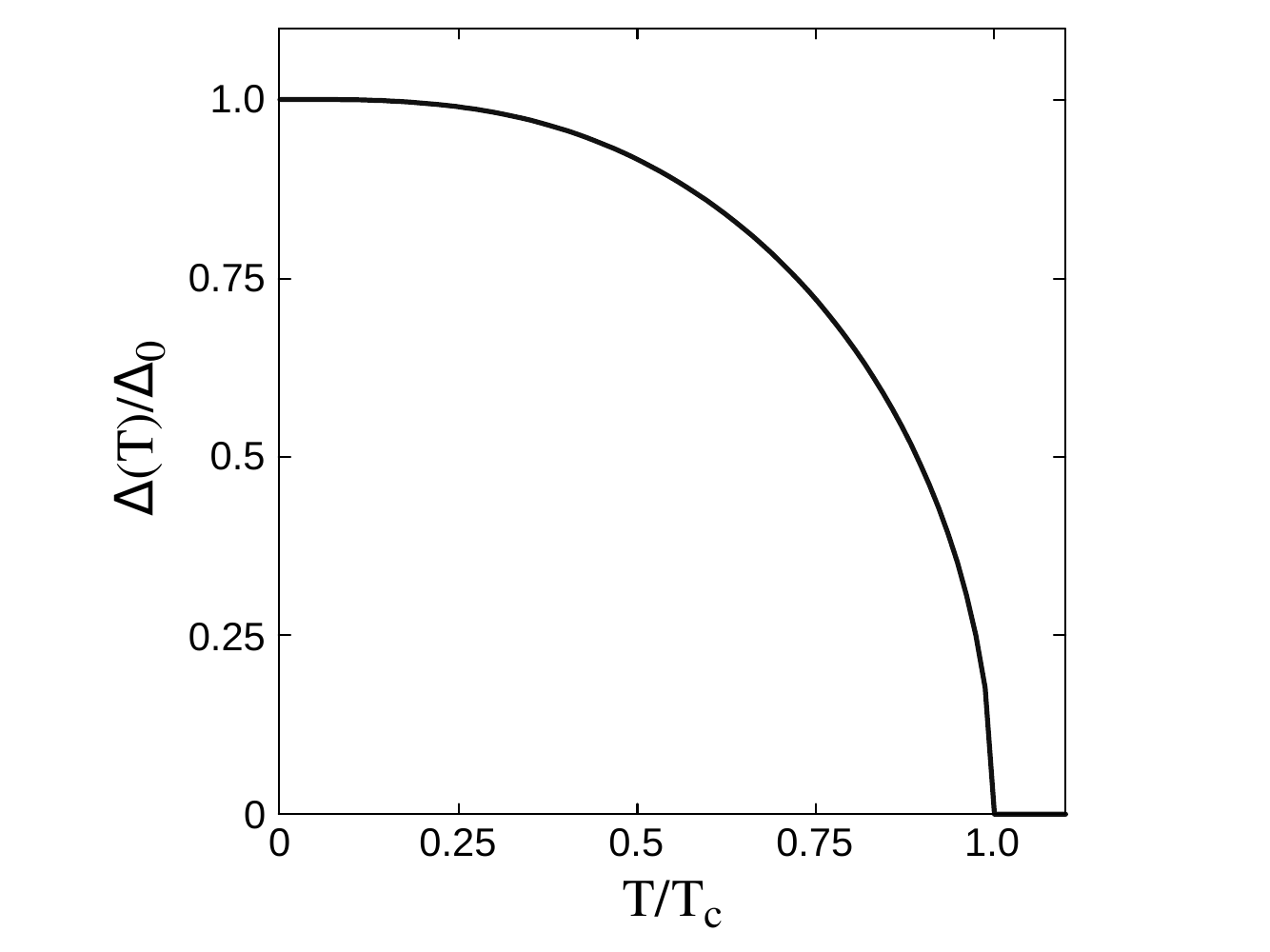}
\caption{Superconducting gap magnitude as a function of the temperature calculated by using Eq. (\ref{bcs}).
The vertical and horizontal axes are normalized by the $\Delta_0 \approx 0.06426$ [eV] and $T_c$, respectively.
}
\label{fig2}
\end{figure}
First, we indicate the calculated temperature dependence of the gap magnitude $\Delta(T)$.
To calculate the BCS gap equation Eq. (\ref{bcs}), we consider only the $\alpha$- and $\beta$-bands that produce $\alpha$- and $\beta$-Fermi surface.
Figure \ref{fig2} shows the normalized gap magnitude $\Delta(T)/\Delta_0$ as a function of the temperature, where $\Delta_0 \approx 0.06426$ [eV] and the scaling factor $A_a$ is set to 0.25 for both the $\alpha$- and $\beta$-bands.
In the following analysis, we use this temperature dependence of the gap magnitude to calculate the $1/T_1$ and PDOS.
\begin{figure}[!t]
\centering
\includegraphics[width = 0.7\linewidth]{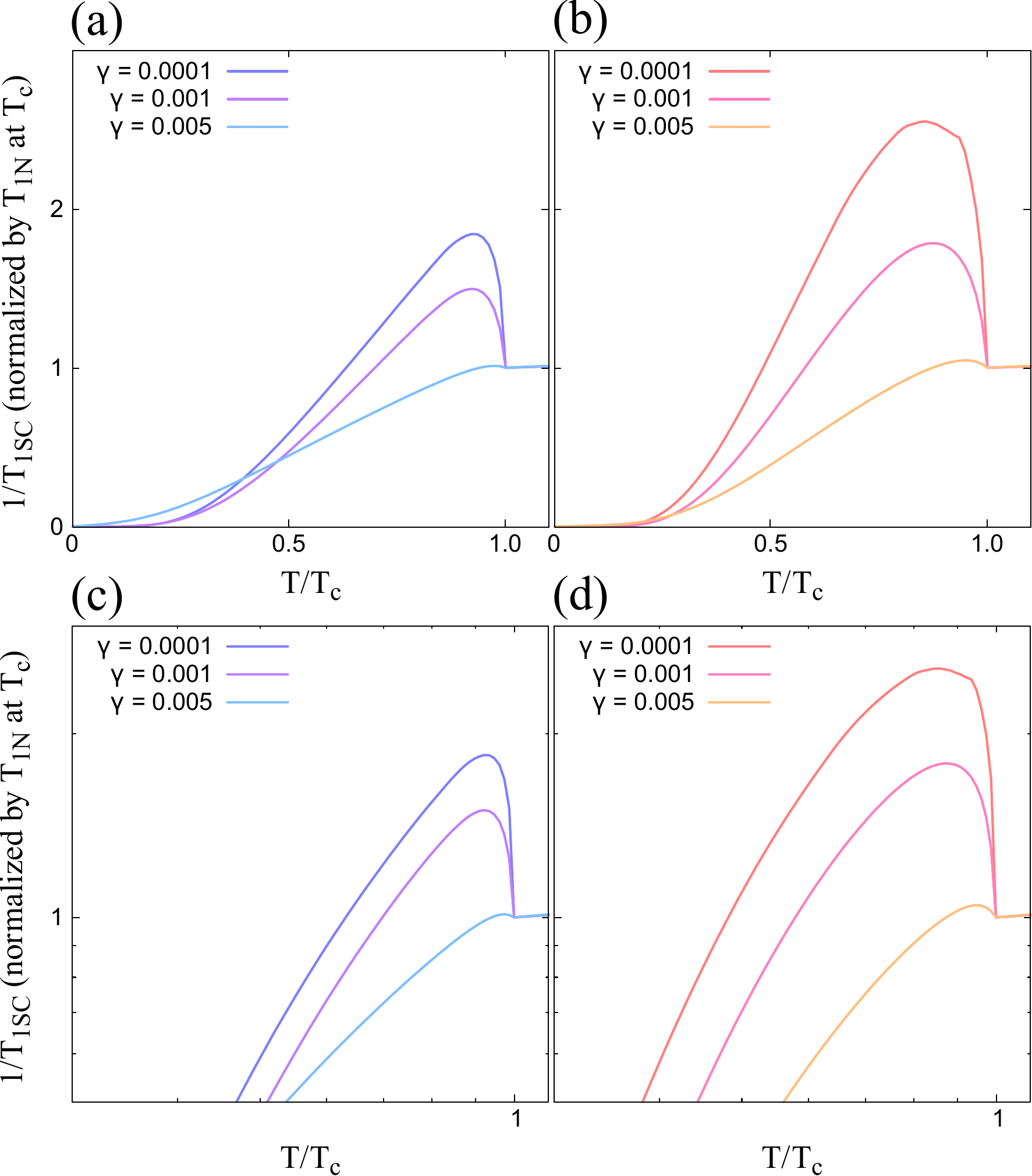}
\caption{
Spin-lattice relaxation rate $1/T_1$ as a function of temperature for $\gamma = 0.0001, 0.001, 0.005$, where the vertical axis is normalized by the normal state value of $T_1$ at $T_c$.
(a) and (b) show $1/T_1$ in the point-node--like gap and isotropic gap cases, respectively.
(c) and (d) show the logarithmic scale plots of (a) and (b), respectively.
}
\label{fig3}
\end{figure}
\begin{figure}[t]
\centering
\includegraphics[width = 0.7\linewidth]{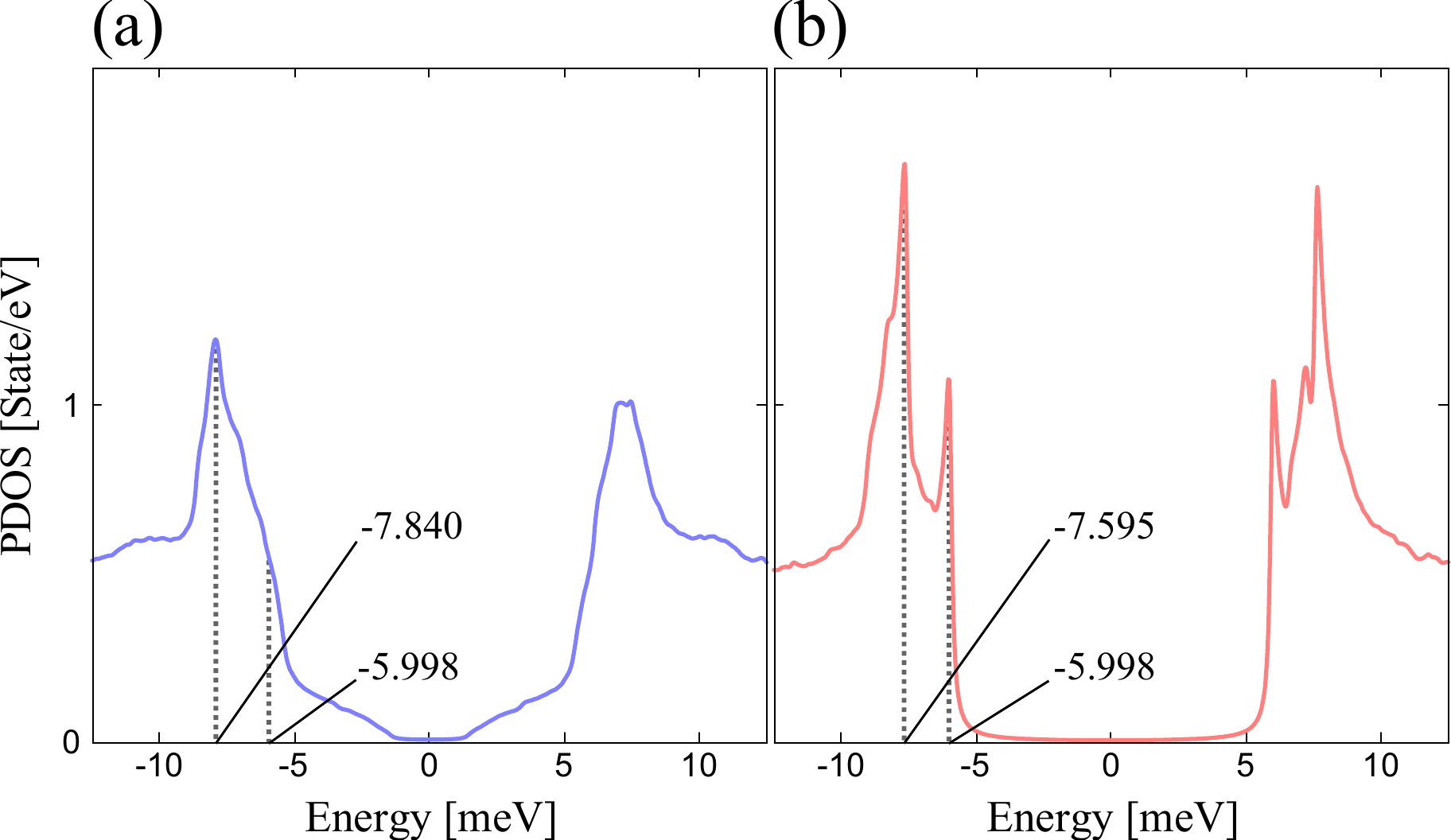}
\caption{
(a) and (b) show the partial density of state (PDOS) as a function of the energy $\varepsilon$ in the point-node--like and isotropic $s$-wave pairing state.
Calculationwas performed under the conditions of $T = 0.9 T_c$ and $\gamma = 0.0001$.
}
\label{fig4}
\end{figure}

Next, we present the numerical results for the spin-lattice relaxation rate $1/T_1$.
For comparison, we display results for both cases the point-node--like and an isotropic gap structures.
Here, the gap function of the isotropic $s$-wave is constracted by $\bar{\Delta}_a = A_a \bar{g}_a \Delta(T)$, where $\bar{g}_a$ is defined as the momentum average of $g_a(\bm{k})$:
\begin{align}
\bar{g}_a \equiv \sqrt{ \frac{1}{N}\sum \limits_{\bm{k}} \left| g_a(\bm{k}) \right|^2 }.
\label{agp}
\end{align}
In this study, the averages of the gap magnitude are $\bar{g}_\alpha \approx 0.8132$ and $\bar{g}_\beta \approx 1.018$.
The temperature dependence of $1/T_1$ for various $\gamma$ is shown in Fig. \ref{fig3}.
In the isotropic $s$-wave pairing situation, the large Hebel-Slichter peak is observed in the case of $\gamma = 0.0001$ and $\gamma = 0.001$ as shown in Fig. \ref{fig3} (b).
Although the Hebel-Slichter peak becomes very small when $\gamma = 0.005$, it can still be observed.
On the other hand, $1/T_1$ as a function of the temperature in the point-node--like gap structure is shown in Fig. \ref{fig3} (a).
Although the Hebel-Slichter peak in the point-node--like gap structure is smaller than one in the isotropic gap structure, it is too large to be neglect.
Similar to the result in the isotropic gap case, the Hebel-Slichter peak is extremely suppressed when $\gamma = 0.005$.
Since these constant parameters $\gamma$ can be related to the damping of quasi-particles due to a system's disorder, the Hebel-Slichter peak in both the point-node--like and isotropic $s$-wave states will be extremely suppressed due to the disorder in the system.
When $\gamma = 0.005$, the Hebel-Slichter peak in the point-node--like $s$-wave pairing becomes negligibly small, however, the Hebel-Slichter peak in the isotropic $s$-wave state also becomes small to the same order of magnitude.
Therefore, although we investigated the possibility that the gap anisotropy leads to the suppression of the Hebel-Slichter peak, it is shown that the accidental nodes of the $s$-wave gap do not explain the absence of the Hebel-Slichter peak in the NMR measurements in UTe$_2$.

Finally, we show the calculated the PDOS of the orbitals $p$ and $p'$ to reveal the reason for the suppression of the Hebel-Slichter peak in the point-node--like $s$-wave case.
Using  Green's function, PDOS is given by
\begin{align}
\rho(\varepsilon) = - \frac{1}{\pi N} \sum \limits_{\bm{k}} \sum \limits_{l = p, p'} \mathrm{Im} G^R_{ll}(\bm{k}, \varepsilon),
\end{align}
where the Green's function is given by Eq. (\ref{gf}).
In this work, the numerical calculation of the PDOS is performed under the following conditions: $T = 0.9T_c$ and $\gamma = 0.0001$, where the gap magnitude $\Delta(T) \approx 0.029595$ [eV].
The plots of the PDOS as a function of the energy $\varepsilon$ in the point-node--like and isotropic $s$-wave pairing states are displayed in Fig. \ref{fig4} (a) and (b), respectively.
Figure \ref{fig4} (b) shows two coherence peaks in the PDOS which are interpreted as follows.
Since in this calculation $\Delta(T) \approx 29.595$ [meV], $\bar{g}_\alpha \approx 0.8132$ and $A_\alpha = 0.25$, the gap scale $|\Delta_\alpha| = |A_\alpha \bar{g}_\alpha \Delta(T)|$ is estimated to be $6.017$ [meV] in the $\alpha$-band.
In the same way, using $\Delta(T) \approx 29.595$ [meV], $\bar{g}_\beta \approx 1.018$ and $A_\beta = 0.25$, the gap scale $|\Delta_\beta| = |A_\beta \bar{g}_\beta \Delta(T)|$ is estimated to be $7.532$ [meV] in the $\beta$-band.
Based on the above estimation, the inner peak and outer peak in Fig. \ref{fig4} (b) correspond to the isotropic superconducting gap of the $\alpha$ and $\beta$-band, respectively.
In Fig. \ref{fig4} (a), however, only one coherence peak PDOS is observed in contrast to the isotropic $s$-wave case.
This peak structure is located at $\varepsilon \approx - 7.840$ [meV], which seems to correspond to the superconducting gap of the $\beta$-band.
The absence of the coherence peak of PDOS around $\varepsilon = -5.998$ is due to the small gap magnitude of the $\alpha$-band superconducting gap, and it is thought that the coherence peak caused by the $\alpha$-band can be buried in the shoulder of the coherence peak from the $\beta$-band.
A comparison Figs. \ref{fig4} (a) and (b) show that the large and sharp coherence peaks are observed in the isotropic $s$-wave case, however, the coherence peaks in the point-node--like gap $s$-wave case are smaller and broader.
This broadening of the PDOS suppresses the low-energy excitation just below $T_c$ and, as a result, the Hebel-Slichter peak in the point-node--like $s$-wave pairing state becomes smaller than that in the isotropic $s$-wave pairing case.

\section{Summary}
In this study, we calculate the spin-lattice relaxation rate $1/T_1$ and the PDOS to explore the behavior of the Hebel-Slichter peak in the multi-band point-node--like $s$-wave superconductivity using the $f$-$d$-$p$ model, which is an effective model of UTe$_2$.
To take into account the degrees of freedom of the sites, the site-resolved calculation of $1/T_1$ in both cases of the point-node--like and isotropic $s$-wave pairing was performed.
This point-node--like $s$-wave pairing was obtained by solving the linearized Eliashberg equation within the TOPT, and it is an unconventional $s$-wave superconductivity.
From these analyses, we discussed the relation between the gap anisotropy and the Hebel-Slichter peak in the model with the degrees of freedom of sites.
The results of our analysis revealed that the Hebel-Slichter peak is robust even in the multi-band point-node--like $s$-wave pairing state, although it is smaller than that in the isotropic $s$-wave pairing state.
In both isotropic and point-node--like $s$-wave cases, the Hebel-Slichter peak becomes smaller as increasing the damping of the quasi-particles $\gamma$.
For a sufficiently large $\gamma$, the Hebel-Slichter peak in the point-node--like $s$-wave state becomes negligibly small.
However, since the Hebel-Slichter peak in the isotropic $s$-wave pairing also becomes small to the same order of magnitude, the scenario that the DOS broadening due to the gap anisotropy prevents the Hebel-Slichter peak is unrealistic.
Therefore, although the point-node--like $s$-wave pairing state agrees with the specific heat measurements, it is inconsistent with the spin-lattice relaxation of the NMR measurements in UTe$_2$.

\vspace{10pt}

\leftline{\textbf{Acknowledgment}}
We wish to thank Mutsuki Iwamoto for helpful discussions and Shin-ichiro Shima for technical assistance of a computer.
We are grateful for the financial support from the Iketani Science and Technology Foundation and JST SPRING, Japan Grant Number JPMJSP2175.


\end{justify}

\end{document}